# HOT QUARK MATTER WITH NEUTRINO CONFINEMENT IN THE FRAMEWORK OF THE LOCAL NAMBU - JONA-LASINIO SU(3) MODEL

G. S. Hajyan[1], G. B. Alaverdyan[2]

*Yerevan State University, Armenia*

The thermodynamic characteristics of hot $\beta$-equilibrium electrically neutral three-flavor quark matter at neutrino confinement are investigated. For the thermodynamic description of such a quark-lepton system, the local SU (3) Nambu-Jona-Lasinio (NJL) model is used, in which also takes into account the 't Hooft interaction, leading to the quark flavor mixing effect. The energy density $\varepsilon$ and pressure $P$ of quark matter are numerically determined for different values of the baryon number density in the range $n_B \in [0.02 \div 1.8]$ fm$^{-3}$ and temperatures in the range $T \in [0 \div 100]$ MeV. The results obtained are compared with the results of cold quark matter calculated within the framework of the same model, but under the assumption that all neutrinos have already left the system. The dependence of the contribution of individual quark flavors to the baryon charge of the system at different temperatures is discussed. Both isothermal and adiabatic speeds of sound in hot quark matter are determined depending on the baryon number density.

## 1. Introduction

Studies of the thermodynamic properties of quark matter, finding the equation of state (EOS) of such a system and clarifying the conditions under which quark deconfinement takes place are an important topic in contemporary physics. The contrast of theoretical results obtained by quantum chromodynamics (QCD) for low values of the baryon chemical potential $\mu_B$ and extremely high temperatures $T$ with experimental data on collisions of heavy relativistic nuclei leads to a deeper understanding of the physical processes taking place in the early universe [1-4]. The region of the QCD phase diagram where the baryon chemical potential $\mu_B$ is fairly high while the temperature is comparatively low corresponds to strange quark stars and hybrid hadron-quark stars, as well as proto-quark stars [5-7]. Comparison of the theoretical predictions in this region with observational data on compact stars makes it possible to enrich our knowledge of the properties of superdense quark matter which could not be studied previously under earthbound laboratory conditions.

---

[1] ghajyan@ysu.am

[2] galaverdyan@ysu.am

In many papers the properties of cold quark matter have been studied in terms of the phenomenological MIT quark bag model and equations of state at zero temperature have been obtained; these are the basis of calculations of the characteristics of hybrid hadron-quark stars, as well as of strange quark stars [8-18].

The NJL model [19,20] has recently often been used to describe quark matter; it was originally proposed for explaining the origin of the nucleon mass taking the spontaneous violation of chiral symmetry into account and was later reformulated for the description of quark matter [21,22]. This model successfully reproduces many features of QCD [23-25]. Combining different modifications of the NJL quark model with different models for describing hadron matter, a number of authors have constructed hybrid EOSs of cold matter and used these to study the properties of neutron stars containing quark matter (e.g., Refs. 26-30).

According to contemporary ideas, a quark star can be formed as the result of a supernova explosion (SNE). Despite colossal progress in computational technology, advances in physics, and the existence of a huge number of computer models of supernova explosions, many questions related to these phenomena are still open.

The process of compressing the central regions of a pre-supernova star to form superdense and superhot (a density of matter greater than nuclear and temperature on the order of $10^{12}$K.) formations lasts no more than a second. Even before these densities and temperatures are reached the material becomes opaque for neutrinos. Thus, the leptonic charge of this new formation will be the same as in the core of the pre-supernova. If this core is of carbon, then the relative leptonic charge $Y_L = 1/2$ (the ratio of the leptonic and baryonic charges of the matter), and in the case of an iron core, $Y_L = 26/56$. This nova formation will be either a proto-neutron star or a proto-neutron star with a quark core. If nature allows the existence of self-coupled quark matter, as indicated by Witten [31], then this will be a proto-quark star.

Questions of thermodynamics, in particular the EOS of hot strange quark matter (HSQM) that is opaque for neutrinos, as well as dense hadron matter in terms of various models, are discussed in Refs. 32-39. HSQM which is opaque to neutrinos is formed in a proto-quark star during a supernova explosion. Hence, these questions are important precisely for the physics of these stars.

The thermodynamic characteristics of HSQM with no neutrinos have been determined in many papers (e.g., Refs. 40-42). These results are of interest for those quark stars in which neutrinos are no longer involved in establishing chemical equilibrium and freely leave the star. These neutrinos are born in HSQM via channels such as bremsstrahlung emission of neutrino pairs by electrons, decay of a plasmon into a neutrino-antineutrino pair, annihilation of an electron-positron pair into a neutrino-antineutrino pair, etc.

This paper is a study of the thermodynamic properties of hot three-flavor quark matter with confinement of neutrinos in terms of a local SU(3) NJL model.

We shall use a "natural" system of units in which the Planck constant, speed of light, and Boltzmann constant are set equal to unity, i.e., $\hbar = c = k_B = 1$.

## 2. Hot quark matter that is opaque to neutrinos in the NJL model

The density of the Lagrangian in terms of the NJL local SU(3) model is given by [23]

$$\mathcal{L}_{NJL} = \bar{\psi}(i\gamma^\mu \partial_\mu - \hat{m}_0)\psi + G \sum_{a=0}^{8}[(\bar{\psi}\lambda_a\psi)^2 + (\bar{\psi}i\gamma_5\lambda_a\psi)^2] - \\ -K\{det_f(\bar{\psi}(1+\gamma_5)\psi) + det_f(\bar{\psi}(1-\gamma_5)\psi)\} \quad (1)$$

Here $\psi$ are the Fermion quark spinor fields $\psi_f^c$ with three flavors $f = u, d, s$ and three colors $c = r, g, b$. The first term is the density of the Lagrangian of free quark fields with a matrix of the masses of the current quarks $\hat{m}_0 = diag(m_{0u}, m_{0d}, m_{0s})$. The second term corresponds to the chiral symmetric four-quark interaction with a coupling constant $G$, where $\lambda_a (a = 1,2,...8)$ are the Gell-Mann matrices and the generators of the SU(3) group in flavor space, $\lambda_0 = \sqrt{2/3}\,\hat{I}$ ($\hat{I}$ is the unit 3x3 matrix). The third term corresponds to the six-quark Kobayashi-Maskawa - 't Hooft interaction [43], which leads to violation of axial $U_A(1)$ symmetry. This interaction is important for splitting of the masses of the $\eta'(958)$ and $\eta(547)$ mesons. Because of this term, in the chiral limit ($m_{0u} = m_{0d} = m_{0s} = 0$) the mass $\eta'$ of a meson increases up to its final value, while the other pseudo-scalar mesons, including $\eta$, remain massless.

Using the mean-field approximation the Lagrangian (1) can be used to obtain an expression for the functional part of the density of the grand thermodynamic potential $\tilde{\Omega}_{NJL}(T, \{M_f\}, \{\mu_f\})$

$$\tilde{\Omega}_{NJL} = -\frac{3}{\pi^2} \sum_{f=u,d,s} \int_0^\Lambda dk\, k^2 E_f(k, M_f) - \\ -\frac{3T}{\pi^2} \sum_{f=u,d,s} \left\{ \int_0^\Lambda dk\, k^2 \left[ \ln\left(1 + e^{-\frac{E_f(k,M_f)-\mu_f}{T}}\right) + \ln\left(1 + e^{-\frac{E_f(k,M_f)+\mu_f}{T}}\right) \right] \right\} + \\ +2G \sum_{f=u,d,s} \left[\sigma_f(T, M_f, \mu_f)^2\right] - 4K\sigma_u(T, M_u, \mu_u)\,\sigma_d(T, M_d, \mu_d)\,\sigma_s(T, M_s, \mu_s), \quad (2)$$

where $\Lambda$ is the momentum at the ultraviolet cutoff which becomes necessary in connection with the unrenormalized character of the NJL model, $E_f(k, M_f) = \sqrt{k^2 + M_f^2}$ is the energy, and $\mu_f$ is the chemical potential of quark-quasiparticles with flavor $f$.

The quark number density is determined by the expression

$$n_f(T, M_f, \mu_f) = \frac{3}{\pi^2} \int_0^\Lambda dk\, k^2 \left[ \frac{1}{1 + e^{\frac{E_f(k,M_f)-\mu_f}{T}}} - \frac{1}{1 + e^{\frac{E_f(k,M_f)+\mu_f}{T}}} \right]. \quad (3)$$

In Eq. (2) for the thermodynamic potential $\widetilde{\Omega}_{NJL}(T,\{M_f\},\{\mu_f\})$, $\sigma_f(T,M_f,\mu_f)$ $(f=u,d,s)$ denotes the quark condensates $\langle \bar{\psi}\psi \rangle$, which are defined as

$$\sigma_f(T,M_f,\mu_f) = \langle \bar{\psi}_f \psi_f \rangle = -\frac{3}{\pi^2} M_f \int_0^\Lambda dk \, \frac{k^2}{E_f(k,M_f)} \left[ 1 - \frac{1}{1+e^{\frac{E_f(k,M_f)-\mu_f}{T}}} - \frac{1}{1+e^{\frac{E_f(k,M_f)+\mu_f}{T}}} \right]. \tag{4}$$

In the mean-field approximation the gap equations for the constituent masses of the quarks $M_u, M_d$, and $M_s$ have the form

$$\begin{aligned} M_u &= m_{0u} - 4G\,\sigma_u + 2K\sigma_d\sigma_s, \\ M_d &= m_{0d} - 4G\,\sigma_d + 2K\sigma_s\sigma_u, \\ M_s &= m_{0s} - 4G\,\sigma_s + 2K\sigma_u\sigma_d. \end{aligned} \tag{5}$$

The density of the grand thermodynamic potential corresponding to the quark component is determined by the expression

$$\Omega_{NJL}(T,\{M_f\},\{\mu_f\}) = \widetilde{\Omega}_{NJL}(T,\{M_f\},\{\mu_f\}) - \widetilde{\Omega}_{NJL}(T=0,\{n_f=0\}). \tag{6}$$

We denote the quark condensates $\sigma_u, \sigma_d$, and $\sigma_s$ at zero temperature $T=0$ and zero densities $n_u = n_d = n_s = 0$ by $\sigma_{u0}, \sigma_{d0}, \sigma_{s0}$, respectively. We denote the masses of the quasiparticle-quarks with flavor $f=u,d,s$ for $n_u = n_d = n_s = 0$ by $M_{f0}$, and the energy, by $E_{f0} = E_f(k, M_{f0})$. Then for the density of the grand thermodynamic potential we obtain

$$\begin{aligned} \Omega_{NJL} = \frac{3}{\pi^2} \sum_{f=u,d,s} \int_0^\Lambda dk\, k^2 \left( E_f(k,M_{f0}) - E_f(k,M_f) \right) - \\ -\frac{3T}{\pi^2} \sum_{f=u,d,s} \left\{ \int_0^\Lambda dk\, k^2 \left[ \ln\left(1+e^{-\frac{E_f(k,M_f)-\mu_f}{T}}\right) + \ln\left(1+e^{-\frac{E_f(k,M_f)+\mu_f}{T}}\right) \right] \right\} \\ +2G(\sigma_u^2 + \sigma_d^2 + \sigma_s^2 - \sigma_{u0}^2 - \sigma_{d0}^2 - \sigma_{s0}^2) - 4K\,(\sigma_u\sigma_d\sigma_s - \sigma_{u0}\sigma_{d0}\sigma_{s0}). \end{aligned} \tag{7}$$

The density of the grand thermodynamic potential of a hot quark-lepton $u,d,s,e,\mu,\nu_e,\nu_\mu,\tau,\nu_\tau$ plasma has the form

$$\begin{aligned} \Omega_{QP} = \frac{3}{\pi^2} \sum_{f=u,d,s} \int_0^\Lambda dk\, k^2 \left( E_f(k,M_{f0}) - E_f(k,M_f) \right) - \\ -\frac{3T}{\pi^2} \sum_{f=u,d,s} \left\{ \int_0^\Lambda dk\, k^2 \left[ \ln\left(1+e^{-\frac{E_f(k,M_f)-\mu_f}{T}}\right) + \ln\left(1+e^{-\frac{E_f(k,M_f)+\mu_f}{T}}\right) \right] \right\} + \\ +2G(\sigma_u^2 + \sigma_d^2 + \sigma_s^2 - \sigma_{u0}^2 - \sigma_{d0}^2 - \sigma_{s0}^2) - 4K\,(\sigma_u\sigma_d\sigma_s - \sigma_{u0}\sigma_{d0}\sigma_{s0}) - \\ -\frac{T}{2\pi^2} \sum_l g_l \int_0^\infty dk\, k^2 \left[ \ln\left(1+e^{-\frac{E_l(k)-\mu_l}{T}}\right) + \ln\left(1+e^{-\frac{E_l(k)+\mu_l}{T}}\right) \right], \end{aligned} \tag{8}$$

where $E_l(k) = \sqrt{k^2 + m_l^2}$, $\mu_l$ is the chemical potential, $m_l$ is the mass of a type $l$ lepton ($l = e, \mu, \nu_e, \nu_\mu, \tau, \nu_\tau,$), and $g_l$ is the degeneracy multiplicity of a type $l$ lepton ($g_e = g_\mu = 2$, $g_{\nu_e} = g_{\nu_\mu} = 1$).

The number density of a type $l$ lepton is determined by the expression

$$n_l(T, \mu_l) = -\frac{\partial \Omega_{QP}}{\partial \mu_l} = \frac{g_l}{2\pi^2} \int_0^\infty dk\, k^2 \left[ \frac{1}{1 + e^{\frac{E_l(k) - \mu_l}{T}}} - \frac{1}{1 + e^{\frac{E_l(k) + \mu_l}{T}}} \right]. \tag{9}$$

For the entropy density of a quark-lepton plasma consisting of particles $f = u, d, s$ and $l = e, \nu_e, \mu, \nu_\mu, \tau, \nu_\tau,$

the NJL model with Eq. (8) yields

$$S_{QP} = -\frac{\partial \Omega_{QP}}{\partial T} =$$

$$= \frac{3}{\pi^2} \sum_{f=u,d,s} \left\{ \int_0^\Lambda dk\, k^2 \left[ \ln\left(1 + e^{-\frac{E_f(k, M_f) - \mu_f}{T}}\right) + \ln\left(1 + e^{-\frac{E_f(k, M_f) + \mu_f}{T}}\right) \right] \right\} +$$

$$+ \frac{3}{\pi^2 T} \sum_{f=u,d,s} \int_0^\Lambda dk\, k^2\, E_f(k, M_f) \left[ \frac{1}{1 + e^{\frac{E_f(k, M_f) - \mu_f}{T}}} + \frac{1}{1 + e^{\frac{E_f(k, M_f) + \mu_f}{T}}} \right] - \frac{1}{T} \sum_{f=u,d,s} \mu_f n_f + \tag{10}$$

$$+ \sum_l \left\{ \frac{g_l}{2\pi^2} \int_0^\infty dk\, k^2 \left[ \ln\left(1 + e^{-\frac{E_l(k) - \mu_l}{T}}\right) + \ln\left(1 + e^{-\frac{E_l(k) + \mu_l}{T}}\right) \right] - \frac{1}{T} \mu_l n_l \right\}.$$

Beginning with the thermodynamic relationship between the energy density and the grand thermodynamic potential $= \Omega + TS + \sum_i \mu_i n_i$, where the sum is taken over all the particle types, we obtain

$$\varepsilon_{QP} = \frac{3}{\pi^2} \sum_{f=u,d,s} \left\{ \int_0^\Lambda dk\, k^2 \left[ E_f(k, M_{f0}) - E_f(k, M_f) \left( 1 - \frac{1}{1 + e^{\frac{E_f(k, M_f) - \mu_f}{T}}} - \frac{1}{1 + e^{\frac{E_f(k, M_f) + \mu_f}{T}}} \right) \right] \right\}$$
$$+$$
$$+ 2G(\sigma_u^2 + \sigma_d^2 + \sigma_s^2 - \sigma_{u0}^2 - \sigma_{d0}^2 - \sigma_{s0}^2) - 4K(\sigma_u \sigma_d \sigma_s - \sigma_{u0} \sigma_{d0} \sigma_{s0}) + \tag{11}$$
$$+ \frac{1}{2\pi^2} \sum_l g_l \int_0^\infty dk\, k^2\, E_l(k) \left[ \frac{1}{1 + e^{\frac{E_l(k) - \mu_l}{T}}} + \frac{1}{1 + e^{\frac{E_l(k) + \mu_l}{T}}} \right].$$

For the pressure we obtain

$$P_{QP} = \frac{3}{\pi^2} \sum_{f=u,d,s} \int_0^\Lambda dk\, k^2 \left(E_f(k,M_f) - E_f(k,M_{f0})\right) +$$

$$+ \frac{3T}{\pi^2} \sum_{f=u,d,s} \left\{ \int_0^\Lambda dk\, k^2 \left[ ln\left(1 + e^{-\frac{E_f(k,M_f)-\mu_f}{T}}\right) + ln\left(1 + e^{-\frac{E_f(k,M_f)+\mu_f}{T}}\right) \right] \right\} - \quad (12)$$

$$-2G(\sigma_u^2 + \sigma_d^2 + \sigma_s^2 - \sigma_{u0}^2 - \sigma_{d0}^2 - \sigma_{s0}^2) + 4K(\sigma_u\sigma_d\sigma_s - \sigma_{u0}\sigma_{d0}\sigma_{s0}) +$$

$$+ \frac{T}{2\pi^2} \sum_l g_l \int_0^\infty dk\, k^2 \left[ ln\left(1 + e^{-\frac{E_l(k)-\mu_l}{T}}\right) + ln\left(1 + e^{-\frac{E_l(k)+\mu_l}{T}}\right) \right].$$

We not that in the formulas for the densities of the grand thermodynamic potential, energy, and entropy, as well as for the pressure, the contributions of a given type of particles and the corresponding antiparticles are combined in a single expression. The expressions for the number density of particles of the $i$-th type, $n_i$, in Eqs. (3) and (9) are actually the differences in the densities of the particles and antiparticles of type $i$.

## 3. The equations of thermodynamic equilibrium for neutrino confinement

Chemical equilibrium of the composition in HSQM is established by mutual conversion processes for the constituents of the substance:

$$\begin{aligned} d &\rightleftarrows u + e^- + \tilde{\nu}_e, & d &\rightleftarrows u + \mu^- + \tilde{\nu}_\mu, \\ s &\rightleftarrows u + e^- + \tilde{\nu}_e, & s &\rightleftarrows u + \mu^- + \tilde{\nu}_\mu, \\ e^- &+ \tilde{\nu}_e \rightleftarrows \mu^- + \tilde{\nu}_\mu. & & \end{aligned} \quad (13)$$

The conditions for chemical equilibrium between quarks and leptons in HSQM with neutrino confinement will be

$$\begin{aligned} \mu_d &= \mu_u + \mu_e - \mu_{\nu_e}, \\ \mu_e - \mu_{\nu_e} &= \mu_\mu - \mu_{\nu_\mu} = \mu_\tau - \mu_{\nu_\tau}, \\ \mu_s &= \mu_u + \mu_e - \mu_{\nu_e}. \end{aligned} \quad (14)$$

These equations should be supplemented by the condition of electrical neutrality and the conservation laws for the baryon and lepton charges.

The electrical neutrality condition has the form

$$\frac{2}{3}n_u - \frac{1}{3}n_d - \frac{1}{3}n_s - n_e - n_\mu - n_\tau = 0. \quad (15)$$

The density of baryon charge is defined by

$$n_B = \frac{1}{3}(n_u + n_d + n_s). \tag{16}$$

The fractions of $e$-lepton, $\mu$-lepton, and $\tau$-lepton charges are specified by the parameters

$$Y_{L_e} = \frac{n_{L_e}}{n_B} = \frac{n_e + n_{\nu_e}}{n_B}, \qquad Y_{L_\mu} = \frac{n_{L_\mu}}{n_B} = \frac{n_\mu + n_{\nu_\mu}}{n_B}, \qquad Y_{L_\tau} = \frac{n_{L_\tau}}{n_B} = \frac{n_\tau + n_{\nu_\tau}}{n_B}. \tag{17}$$

If the independent input variables are taken to be the temperature $T$, baryon number density $n_B$, lepton charge densities $n_{L_e}$, $n_{L_\mu}$, and $n_{L_\tau}$, , then, neglecting neutrino oscillations, Eqs. 3-5, 9, and 14-17 can he used to determine the densities and chemical potentials of all the HSQM particles, as well as the effective quark masses $M_u$, $M_d$, and $M_s$, and the quark condensates $\sigma_u$, $\sigma_d$, $\sigma_s$.

In this paper the range of the baryon charge density of the HSQM is limited above by $n_B \approx 10 n_0$. It is precisely this region which is of physical interest for the physics of hybrid hadron-quark stars, as well as strange quark stars. In this sort of HSQM, besides deconfined $u, d,$ and $s$ quarks there are electrons, muons, and all possible types of neutrino, along with the antiparticles corresponding to them. Because of the high mass of the tau-lepton and comparatively low value of the chemical potential of electrons and muons, for the range of baryon charge we are examining there will be no tau-leptons in the HSQM, while tau-neutrinos $\nu_\tau$, will be formed through decay of a plasmon or as a result of neutrino oscillations.

In each elementary act of interaction between leptons and baryons or with quarks, the corresponding lepton charge is conserved. We discuss an HSQM that was formed during the implosion of the central regions of a pre-supernova star without loss of lepton charges. Since only electron leptons were present in the material prior to the onset of the implosion, in the newly formed HSQM only the electron lepton charge will be nonzero. It should be noted that neutrino oscillations can disrupt this relationship of the lepton charges. Muons can show up both before and after breakup of baryons.

Neglecting effects owing to neutrino oscillations, we may assume that $n_{L_\tau} = 0$, i.e., the densities of $\nu_\tau$, and $\tilde{\nu}_\tau$, are the same and are determined exclusively by the temperature of the HSQM, while the chemical potential of these neutrinos equals zero.

Neutrino oscillations associated with the existence of mass in neutrinos were predicted more than half a century ago by Pontecorvo [44] and discovered experimentally in 2015. If neutrino oscillations in the HSQM of a proto-quark star take place so rapidly that the densities and chemical potentials of all the types of neutrino become the same, i.e.,

$$n_{\nu_e} = n_{\nu_\mu} = n_{\nu_\tau} = n_{\nu 0},$$
$$\mu_{\nu_e} = \mu_{\nu_\mu} = \mu_{\nu_\tau} = \mu_\nu, \tag{18}$$
$$n_\nu = 3 n_{\nu 0},$$

then when neutrino oscillations are taken into account, the three conservation equations for the lepton charges of the different types (17) are replaced by a single conservation equation for the total lepton charge:

$$n_L = n_e + n_\mu + n_\nu. \tag{19}$$

For the HSQM thermodynamics in this case all the types of neutrino are formally indistinguishable. Their total density $n_\nu$, temperature $T$, and chemical potential $\mu_\nu$ are related by Eq. (9) with a "degeneracy" $g_\nu = 3$. The equality of all the neutrino chemical potentials according to Eq. (14) also implies equality of the electron and muon chemical potentials. In this way, when neutrino oscillations are taken into account the number of equations for the thermodynamic equilibrium is reduced by two equations.

We recall that in all the formulas given here the number density $n_i$ of particles of type $i$ is actually the difference between the densities of the particles and antiparticles of that type.

## 4. Results of numerical calculations

Numerical calculations were carried out for this paper with values for the parameters of the NJL model given in Ref. 23:

$$m_{0u} = m_{0d} = 5.5 \text{ МэВ}, \quad m_{0s} = 140.7 \text{ МэВ}, \quad \Lambda = 602.3 \text{ МэВ},$$
$$G = 1.835/\Lambda^2, \quad K = 12.36/\Lambda^5.$$

For the fractions of lepton charges. i.e., the lepton charges per unit baryon charge, the values $Y_{L_e} = 0.4$, $Y_{L_\mu} = 0$, and $Y_{L_\tau} = 0$ were used.

As noted above, in the matter of the central regions of a pre-supernova star prior to the implosion the relative lepton charge $Y_e = n_e/n_B$ (the number of electrons per unit baryon charge) lies in the range of $26/56 \div 1/2$. In our calculations it is assumed that $Y_e = 0.4 < 26/56$, so as to somehow compensate the partial reduction in the lepton charge during the implosion before formation of the proto-quark star.

Effects associated with neutrino oscillations are neglected in this paper. In this case only electron and muon neutrinos are involved in establishing the thermodynamic equilibrium.

For the specified temperatures $T$, baryon charge densities $n_B$, and relative lepton charges $Y_{L_e}, Y_{L_\mu}$, and $Y_{L_\tau}$ solution of the system of Eqs. 3-5, 9, and 14-17 makes it possible to determine the constituent masses of the quarks $M_u$, $M_d$, and $M_s$, and the quark condensates $\sigma_u$, $\sigma_d$, $\sigma_s$, as well as the densities and chemical potentials of all the constituents of the HSQM. Finding these characteristics makes it possible in turn to calculate the remaining thermodynamic quantities of the HSQM: the energy density and entropy density, as well as the pressure.

Since the lepton charge of the $\tau$-neutrino is zero in the case of interest to us, in accordance with Eq. (9), independently of the found solution, the energy density and pressure of the $\tau$-neutrinos are given by

$$\varepsilon_{\nu_\tau} = 3p_{\nu_\tau} = \frac{7}{240}\pi^2 T^4 \qquad (20)$$

Figure 1 shows the isothermal dependences of the pressure $P$ of the HSQM on the baryon charge density $n_B$, for $T = 0, 40, 60, 80,$ and 100 MeV. In this figure two curves are shown for the cold state (T = 0). The smooth curve $P^{(0)}(n_B) = P(n_B, T = 0)$ corresponds to cold strange quark matter (CSQM) with neutrinos present at a lepton charge $Y_{L_e} = 0.4$. Of course, this sort of state cannot be realized since in fully degenerate matter the mean free path for neutrinos is infinite and they freely escape the matter. Ultimately, a state with $Y_{L_e} \sim 10^{-3}$ is established. This curve is shown as the limiting position of the $P = P(n_B, T)$ curves with $Y_{L_e} = 0.4$ as $T \to 0$. As a proto-quark star cools with loss of neutrinos, i.e., as the lepton charge decreases, the EOS of the HSQM of the star asymptotically approaches $P_0(n_B) = P(n_B, n_\nu = 0, T = 0)$ [30] without neutrino content (the dashed curve in Fig. 1). The shift of these two $P^{(0)}(n_B) = P(n_B, T = 0)$ curves is caused specifically by the large difference between the lepton charges in these two states of the CSQM. For $n_{0B} \approx 0.38$ fm$^{-3}$ the pressure of the CSQM equals zero. The CSQM does not exist below this density. An HSQM with a temperature below T $\approx$ 20 MeV is the same. Above this temperature up to T $\approx$ 50 MeV the pressure of an HSQM is always positive, but at the point $n_B \approx 0.3$ fm$^{-3}$ the $P = P(n_B, T)$ curves have a minimum. Thus, for $n_B \leq 0.3$ fm$^{-3}$ at temperatures $20 \leq T \leq 60$ MeV, an HSQM is also not realized.

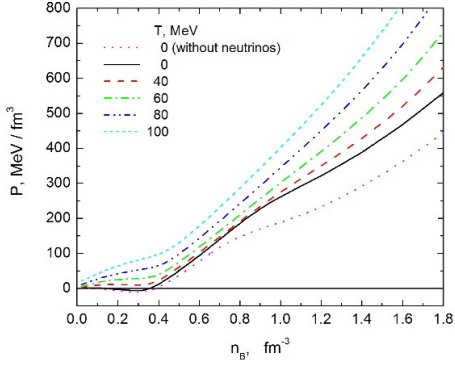
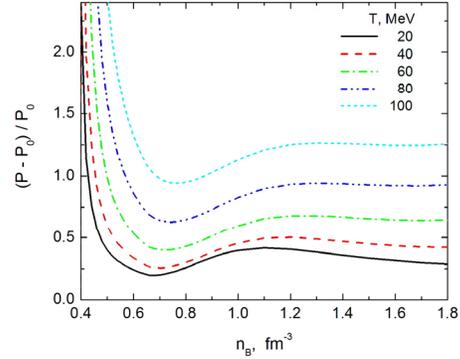

Fig. 1. Pressure $P$ as a function of the baryon number density $n_B$ for strange quark matter at different temperatures $T$. The smooth curve corresponds to cold matter with neutrino confinement and the dotted curve, to cold matter without neutrino confinement [30].

Fig. 2. Relative variation in the pressure $(P - P_0)/P_0$ ($P_0 = P(n_B, n_\nu = 0, T = 0)$) of strange quark matter with relative lepton charge $Y_{L_e} = 0.4$ with complete cooling and departure of all neutrinos in the dependence on the baryon charge density $n_B$ for different values of the initial. temperature.

In Fig. 2 isothermal dependences of the relative change in the pressure $(P - P_0)/P_0$ ($P_0 = P(n_B, n_\nu = 0, T = 0)$) at initial temperatures of T = { 20; 40; 60; 80; 100} MeV are shown for complete cooling and the escape of all neutrinos at baryon charge density $n_B$. The large values of this ratio near $n_{0B} \approx 0.38$ fm$^{-3}$ are caused by the zero pressure $P_0$.

For the physics of superdense celestial objects the range of densities $n_B > (3 \div 4)n_0$, where $n_0 = 0.16$ fm$^{-3}$ is the nuclear saturation density, is of interest. In this region the HSQM pressure at $T = 100$ MeV is greater by more than a factor of two than the CSQM pressure. This means that during cooling, a proto-quark star will be strongly compressed, which leads to the release of a huge amount of gravitational energy.

Figure 3 shows the isothermal pressure $P$ of the HSQM as a function of the energy density $\varepsilon$ for the same values of the temperature. The smooth and dashed curves are the same as in Fig. 1.

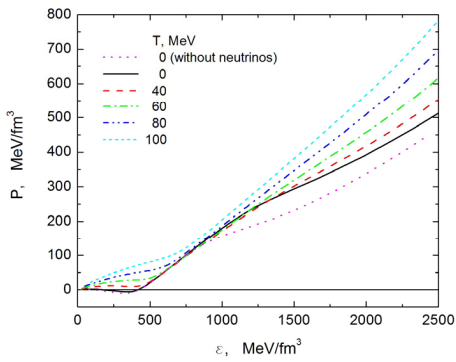
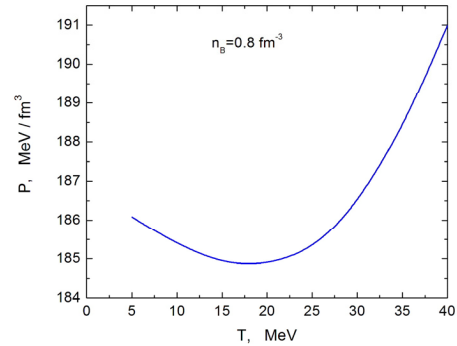

Fig. 3. Pressure $P$ as a function of energy density $\varepsilon$ for strange quark matter at different temperatures $T$. The smooth curve corresponds to cold matter with neutrino confinement and the dotted curve, to cold matter without neutrino confinement.

Fig. 4. Pressure $P$ as a function of temperature $T$ for HSQM with a fixed value of the baryon charge density, $n_B = 0.8$ fm$^{-3}$.

The characters of the $P = P(\varepsilon, T)$ and $P = P(n_B, T)$ curves are mostly the same. Only in the region $700 \leq \varepsilon \leq 1250$ MeV/fm$^3$ and T $\leq 60$ MeV for equal energy densities is the HSQM pressure less than the pressure of the CSQM. For a given energy density the baryon charge density in the CSQM is greater than in the HSQM and it seemed that this could explain this behavior of these curves. The fact that only with this it is not possible to explain this behavior of the $P = P(\varepsilon, T)$ curves is evident from Fig. 4, which shows the temperature dependence of the HSQM pressure for a constant baryon charge density of $n_B = 0.8$ fm$^{-3}$.

The HSQM pressure falls with increasing temperature to a minimum value and then rises. This should not seem strange if we recall the properties of water at normal pressure and t$< 4^0 C$; with increasing pressure the temperature of water falls owing to the way the particles that form the system interact. The above mentioned behavior of the pressure of HSQM at relatively low temperatures (Figs. 3 and 4) can be explained by the formation of an $s$ quark and the contribution of this quark with the other particles to the interaction energy of this quark. According to the NIL model, excess formation of a strange quark in HSQM sets in when $n_B = 0.7 \div 0.8$ fm$^{-3}$. In accordance with the Le Chatelier principle the creation of particles of a new type reduces the pressure rise. For low temperatures this shows up clearly in Fig. 1.

Figure 5 shows the relative contributions of the individual quark flavors $\frac{1}{3}n_f/n_B$, $(f = u, d, s)$ to the baryon charge of the system as a function of the baryon charge density $n_B$ for different temperatures $T = 0, 20,$ and $100$ MeV.

In CSQM the threshold for creation of a strange quark has a strictly determined value: $n_B \approx 0.66$ fm$^{-3}$. For T $\neq$ 0 because of the high-energy "tails" a Fermi distribution of the particles in an HSQM of $s$ quarks is present for an arbitrary density. If the effective threshold for creation of an $s$ quark can be regarded as $n_B = 0.5 \div 0.6$ fm$^{-3}$ at $T = 20$ MeV, then for $T = 100$ MeV we cannot speak of such a threshold. This is clearly shown in Fig. 5. As expected, cooling of an HSQM with a constant baryon charge density $n_B$ is accompanied by a monotonic reduction in the energy density.

Figure 6 shows isothermal dependences of the energy density on the baryon charge density $n_B$ for temperatures $T = 40, 60, 80,$ and $100$ MeV. In the limit of $n_B \to 0$, where there are no more quarks, the "substance" only contains lepton-antilepton pairs. Thus, the energy density in this limit has a nonzero value.

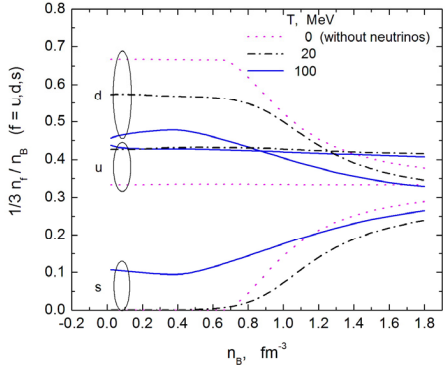

Fig. 5. Individual contributions of different quark flavors to the baryon charge of a $n_f/(3n_B)$, $f = u, d, s$ system as functions of the baryon charge density $n_B$ for strange quark matter at different temperatures $T$.

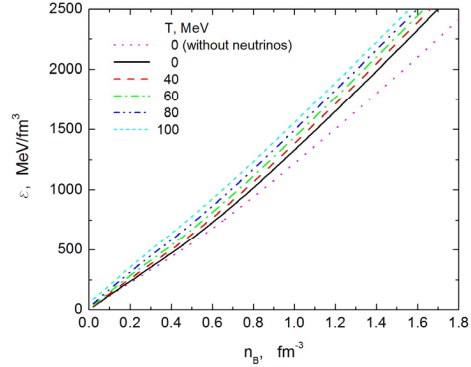

Fig. 6. The energy density as a function of the baryon charge density $n_B$, for strange quark matter at different temperatures $T$. The smooth curve corresponds to cold matter with confinement of neutrinos and the dotted curve, to cold matter without confinement of neutrinos.

Figure 7 shows the isothermal dependences for initial temperatures of $T = 20, 40, 60, 80,$ and $100$ MeV On the baryonic charge density $n_B$ of the relative variation in the energy density $(\varepsilon - \varepsilon_0)/\varepsilon_0$ ( $\varepsilon_0 = \varepsilon(n_B, n_v = 0, T = 0)$ ) for complete cooling and the departure of all neutrinos.

In the limit $n_B \to \infty$ the quark component approaches an asymptotically free state: it becomes a degenerate ultrarelativistic ideal gas. In the cold state, in such a gas the densities of all quarks are the same and there are no electrons. In this limit for an HSQM with $L \neq 0$ (leptons exist in the HSQM) the densities of the individual quarks are not equal. Thus, the energy in this state differs from the CSQM energy. In the limit $n_B \to$

∞ the ratio $(\varepsilon - \varepsilon_0)/\varepsilon_0$ will not depend on temperature. In Fig. 7 the maximum value of the baryon charge density is not so high, but the above-mentioned behavior is already noticeable.

The numerical value of $(P - P_0)/P_0$ in HSQM exceeds the value of $(\varepsilon - \varepsilon_0)/\varepsilon_0$ by a factor of up to $3 \div 5$. Given that the energy density and pressure of a superdense substance are almost mutually proportional quantities, this seems strange. It is easy to confirm that this difference between these quantities is related to the presence of a zero pressure in the CSQM for a nonzero baryon charge density $n_B = n_{0B} \approx 0.38$ fm$^{-3}$.

It is clear from Fig. 7 that, as it cools at a constant baryon number density, an HSQM with an initial temperature of $T = 100$ MeV will lose (mainly as a neutrino flux) up to 30% of its energy. As a proto-quark star cools it will be compressed with simultaneous release of a huge amount of energy. Owing to these two processes the mass of a proto-quark star will decrease significantly. Calculations with the MIT bag model show that the mass of a prom-quark star during cooling may decrease to 25% [45]. The appropriate calculations will show how the mass of a proto-quark star will decrease during cooling according to the NJL model.

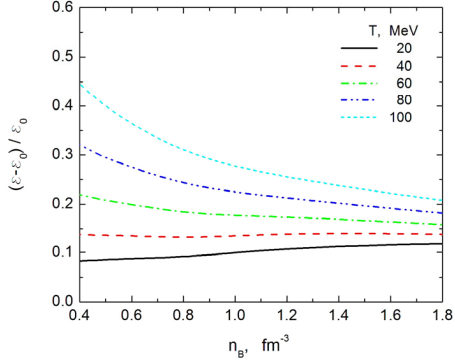
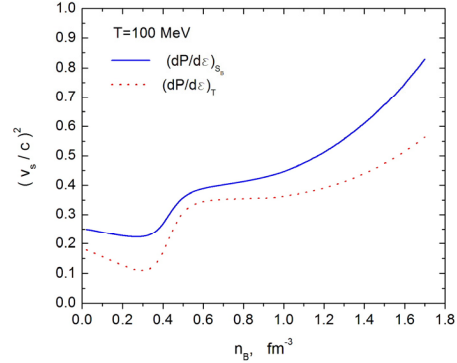

Fig. 7. Relative variation in the energy density $(\varepsilon - \varepsilon_0)/\varepsilon_0$ ( $\varepsilon_0 = \varepsilon(n_B, n_\nu = 0, T = 0)$ ) of strange quark matter with a relative lepton charge of $Y_{L_e} = 0.4$ for complete cooling and departure of all neutrinos with the baryon charge density $n_B$ for different initial temperatures $T$.

Fig. 8. The squares of the adiabatic $v_s^{(S_B)}$ and isothermal $v_s^{(T)}$ sound speeds in HSQM for a temperature $T = 100$ MeV as functions of the baryon charge density $n_B$ neglecting the finiteness of the rates of the relaxation processes leading to changes in the composition of the substance.

The propagation velocity of mechanical perturbations in a material medium is referred to as the speed of sound $v_s$ in this medium. In gases and liquids this speed is defined both by the equation of state and the character of the processes during the propagation of sound. In ordinary gases the periodic compression and rarefaction is accompanied by a rise and fall of temperature. For frequencies $f < 10$ Hz thermal conductivity is able to keep the process isothermal, while for $f < 20$ Hz the process proceeds without heat transfer, i.e., the oscillations propagate adiabatically. Thus, isothermal $v_s^{(T)} = c\sqrt{(dP/d\varepsilon)_T}$ and adiabatic $v_s^{(S_B)} = c\sqrt{(dP/d\varepsilon)_{S_B}}$ sound speeds are distinguished. We bring in this material from the physics textbooks in order to illustrate the source of an error in defining the speed of sound in both cold degenerate and hot superdense matter.

In many papers on the standard equation of state $P = P(\varepsilon, T)$ of superdense matter, by analogy with a Boltzmann gas the sound speed is defined as

$$v_s^{(S_B)} = c\sqrt{(dP/d\varepsilon)_{S_B}} \,. \tag{21}$$

However, if structural changes occur as the sound propagates in the medium, then the domain of validity of Eq. (21) has to be refined.

Figure 8 shows the squares of the adiabatic and isothermal sound speeds in HSQM with a temperature of $T = 100$ MeV as functions of the baryon charge density under the assumption that at all times the composition of the particles with a periodic density oscillation corresponds to the equilibrium state. In the adiabatic approximation the speed of sound is determined for different equations of state of superdense matter without accounting for relaxation phenomena in Refs. 38, and 46-48.

An HSQM is a relaxing system. A sound wave may destroy the thermodynamic equilibrium between components of the HSQM, which has a several characteristic relaxation times. One is determined by the magnitude of the coefficient of thermal conductivity and others, by the characteristic times of the fundamental interactions. For HSQM the numerical values of these times differ greatly. Accounting for the influence of all the possible relaxation processes on the sound speed in HSQM requires separate study, which goes beyond the scope of this paper. Of course, the speed of a sound wave in HSQM will depend on the frequency $\omega_s$. A sound wave with this frequency in HSQM will have the highest speed if the composition is fixed when the wave propagates, i.e., the acoustic oscillations of the medium proceed so rapidly that the material is unable to readjust. If, on the other hand, the period of the density oscillations in the sound wave is so high that at each moment the HSQM is in equilibrium (the medium is able to readjust), then the sound speed will be minimal. This corresponds to the lowest curve in Fig. 8.

Given that the lifetime of a free neutron is on the order of fifteen minutes, the slowest relaxation process is the readjustment of the quark component via reactions (13). Superdense matter has a huge thermal conductivity. Thus, a sound wave in HSQM will be an isothermal process over a broad range of frequencies without a change in the quark composition. A detailed study of the frequency dependence of the sound speed in HSQM is a topic for separate study.

## 5. Conclusion and critical comments

The thermodynamic characteristics of hot quark matter that is opaque to neutrinos have been determined in terms of the NJL model neglecting neutrino oscillations and the vector and vector-axial channels for interactions among quarks. In our calculations three sorts of neutrino have been taken into account consistent with the standard model of elementary particle theory.

As opposed to the MIT quark bag model, the NJL model does not contain a self-coupled state of quark matter, so in this paper the term "proto-quark star" applies only to stars with masses close to the maximum possible value, for which the matter is mainly in the quark state. True, a proto-quark star state may exist with neutrino confinement for some time, a few minutes or even longer, but this is an important stage in the life of a supernova explosion If a neutrino is not confined during the implosion of the central regions of a pre-supernova star, then the energy of the individual neutrino will be an order of magnitude lower than when they are confined. Although the overall amount of energy carried away by neutrinos in both cases is roughly the same, because of a comparatively larger cross section for interactions with the matter in the surrounding proto-quark star, the more energetic neutrinos may play a decisive role in the further fate of the SNE.

Knowledge of the physical properties of quark matter in the range of 3-5 times nuclear density is important for the astrophysics of superdense celestial objects. Theoretical studies of the properties of quark matter in the region below these densities are of interest for understanding and explaining experimental data on collisions of high-energy atomic nuclei.

This work was carried out at the scientific-research laboratory for the physics of superdense stars in the Department of applied electrodynamics and modeling at Erevan State University and financed by the committee on science of the Ministry of education, science, culture, and sport of the Republic of Armenia.